\documentclass[12pt]{article}
\usepackage{graphicx}

\newcommand\pubdate{August, 2015}

\def\asu{Department of Physics\\
  Arizona State University, Tempe, AZ 85287-1504, USA}
\def\support{\footnote{Work supported by the National Science
    Foundation under Grant Numbers PHY-1068286 and PHY-1403891.}}

\def\Title#1{\begin{center} {\Large #1 } \end{center}}
\def\Author#1{\begin{center}{ \sc #1} \end{center}}
\def\Address#1{\begin{center}{ \it #1} \end{center}}

\newcommand\pubblock{\rightline{\begin{tabular}{l} 
         \pubdate  \end{tabular}}}
\newenvironment{Abstract}{\begin{quotation}  }{\end{quotation}}
\newenvironment{Presented}{\begin{quotation} \begin{center} 
             PRESENTED AT\end{center}\bigskip 
      \begin{center}\begin{large}}{\end{large}\end{center} \end{quotation}}
\def\Acknowledgements{\bigskip  \bigskip \begin{center} \begin{large}
             \bf ACKNOWLEDGEMENTS \end{large}\end{center}}

\def\beq{\begin{equation}}
\def\eeq#1{\label{#1}\end{equation}}
\def\eeqn{\end{equation}}

\def\beqa{\begin{eqnarray}}
\def\eeqa#1{\label{#1}\end{eqnarray}}
\def\eeqan{\end{eqnarray}}

\let\bar=\overbar

\def\Dslash{\not{\hbox{\kern-4pt $D$}}}
\def\dslash{\not{\hbox{\kern-2pt $\del$}}}

\def\msb{{\bar{\ssstyle M \kern -1pt S}}}

\begin{document}
\begin{titlepage}
\pubblock

\vfill
\Title{A New Dynamical Picture for the Production and Decay of the $XY
  \! Z$ Mesons}
\vfill
\Author{Richard F. Lebed\support}
\Address{\asu}
\vfill
\begin{Abstract}
  I introduce an entirely new dynamical description for exotic
  charmoniumlike hadrons, based upon the competing effects of the
  strong attraction between quarks in a diquark, and the inability of
  the diquark to hadronize on its own due to being a color nonsinglet.
  This mechanism naturally explains, for example, the strong
  preference of the $Z(4475)$ to decay to $\psi(2S)$ rather than the
  $J/\psi$, the existence of a state $X(4630)$ that decays to
  $\Lambda_c$ baryon pairs, and why some but not all exotics lie near
  hadronic thresholds.  Owing to high-energy constituent counting
  rules, the four-quark nature of the states produces major changes to
  both the high-$s$ scaling of cross sections for producing such
  states and to the potency of the cusp effect of attracting
  resonances to pair-production thresholds.  The recently observed
  $P_c^+$ pentaquark candidates are seen to fit naturally into this
  scheme.
\end{Abstract}
\vfill
\begin{Presented}
The 7th International Workshop on Charm Physics (CHARM 2015)\\
Detroit, MI, 18-22 May, 2015
\end{Presented}
\vfill
\end{titlepage}
\def\thefootnote{\fnsymbol{footnote}}
\setcounter{footnote}{0}
%

\section{Introduction}

In addressing an audience that specializes in charm physics, it is
hardly necessary to emphasize the importance of the discovery of
``exotic charmoniumlike states'' such as $X(3872)$ and $Z(4475)$.
Evidence continues to mount that they are tetraquark $c\bar c \, q_{1
\,} \! q_2$ states, the first hadrons not classifiable under the
$q\bar q$ meson/$qqq$ baryon scheme.  Approximately 20 exotic
charmoniumlike states, at various levels of confirmation, have been
observed at BaBar, Belle, BESIII, CDF, CLEO, CMS, D$\emptyset$, and
LHCb (for recent reviews, see
Refs.~\cite{Brambilla:2014jmp,Esposito:2014rxa}).  The situation has
become even more interesting with the LHCb
observation~\cite{Aaij:2015tga} of pentaquark $c\bar c u u d$
candidates $P_c^+ (4380)$ and $P_c^+ (4450)$.

The nonobservation of unambiguous QCD exotics until only a few years
ago is one of the more perplexing facts in the history of hadronic
physics.  QCD allows for the formation of many more color-singlet
combinations than just those in the $q\bar q$/$qqq$ paradigm:
glueballs, hybrids, tetraquarks, pentaquarks, and so on.  Presumably,
understanding the way in which these new states are assembled from
their quark components will provide key evidence for which phenomena
are allowed by QCD and which are not.

In the case of the tetraquarks, a variety of different physical
pictures have been proposed.  When only the neutral $X$ and $Y$ states
were known, it was possible to suppose they were all either
conventional $c \bar c$ charmonium at unexpected masses or hybrid $c
\bar c g$ ($g$ = valence gluon) states; but the discovery of the
charged $Z$ states mandated a 4-quark structure, at least for those
states.  However, as time progressed, hadronic transitions between the
$X$, $Y$, and $Z$ states were observed, suggesting a common structure.
Moreover, hybrids are constructed easily only in certain $J^{PC}$
channels such as $1^{++}$, while exotics in various $J^{PC}$ channels
have now been observed.

The remarkable proximity of some of these states to 2-meson thresholds
has spawned three major physical pictures to describe them.  Consider,
{\it e.g.}, $X(3872)$:
\begin{eqnarray}
m_{X(3872)} - m_{D^{*0}} - m_{D^0} & = & -0.11 \pm 0.21 \, {\rm MeV}
\, , \nonumber \\
m_{X(3872)} - m_{J/\psi} - m_{\rho^0_{\rm peak}} & = & -0.49 \pm 0.30
\, {\rm MeV} \, , \nonumber \\
m_{X(3872)} - m_{J/\psi} - m_{\omega_{\rm peak}} & = & -7.88 \pm 0.21
\, {\rm
  MeV} \, . \label{eq:massdiffs}
\end{eqnarray}
In comparison, the deuteron, which is considered a loosely bound $p \,
n$ state, has a binding energy of $2.22$~MeV, some {\em twenty
times\/} larger than the central value in the first of
Eq.~(\ref{eq:massdiffs}).  One natural explanation for these numerical
oddities is that the tetraquarks are molecules of two (color-singlet)
mesons, held together by residual ``color van der Waals'' forces.  In
this picture, the binding is accomplished by the exchange of light
mesons, particularly pions.  This {\it meson molecule\/} picture is
the most popular of all paradigms for the tetraquarks, having been
studied in hundreds of papers.  However, the proximity of the
charmonium-unflavored meson threshold for many of the exotics also
suggests the possibility of {\it
hadrocharmonium}~\cite{Dubynskiy:2008mq}, a picture in which a compact
charmonium state lies embedded in a light-quark hadronic cloud, and
retains much of its identity until the decay.  The values in
Eq.~(\ref{eq:massdiffs}) also suggest the possibility that at least
some of the exotics are not true resonant states, but rather an effect
caused by the rapid opening of meson-meson thresholds, which creates a
peak in the production rate near the threshold resembling that due to
a true resonance; this phenomenon is the so-called {\it cusp\/} or
{\it threshold\/} effect, which has been known for decades in
light-quark systems, but was first applied to the charm sector in
Ref.~\cite{Bugg:2008wu}.

In this talk, however, I wish to advocate for yet one more physical
picture for the tetraquarks, which is based upon a well-known yet
under-appreciated feature of QCD\@.  The strongest attraction between
two color-fundamental quarks is that between a color-{\bf 3} quark and
a color-$\bar {\bf 3}$ antiquark into a color singlet, which is of
course the basis of the color structure of a meson.  However, it is
not the only attractive channel: Two {\em quarks\/} can combine into
an attractive color-$\bar {\bf 3}$ combination, and the strength of
this attraction at short distances is fully half as large as that of
the singlet channel.  One therefore expects, at least in some physical
circumstances, the formation of fairly compact {\em diquark\/} states
within hadronic systems.

In a system with two quarks and two antiquarks, one therefore has two
natural ways to assemble the state: Either one has two associated
color-singlet quark-antiquark pairs, as in the molecular or
hadrocharmonium pictures, or one pairs the quarks into a diquark, and
the antiquarks into an antidiquark.  This {\it diquark picture\/} for
exotic charmoniumlike states was first studied in
Ref.~\cite{Maiani:2004vq}, and was greatly improved to reflect the
results of more recent experiments in Ref.~\cite{Maiani:2014aja}.  The
greatest difficulties with this picture and the others listed above
are summarized in Sec.~\ref{sec:Limitations}.

Since Refs.~\cite{Maiani:2004vq,Maiani:2014aja} discuss the tetraquark
states in terms of spin structure in a Hamiltonian formalism, they
implicitly treat the tetraquark as a diquark-antidiquark molecule.
The specific diquark picture to be described here, introduced in
Ref.~\cite{Brodsky:2014xia}, treats the tetraquark as a type of bound
state not previously discussed: The diquark and antidiquark do not
actually orbit one another, but remain bound together solely through
color confinement.  The means by which such states form and decay is
discussed in Sec.~\ref{sec:BHLpicture}.  The use of this picture to
probe the multiquark nature of states and its combination with the
cusp effect, as well as to provide an explanation of the new $P_c^+$
states, is briefly discussed in Sec.~\ref{sec:Applications}.
Conclusions appear in Sec.~\ref{sec:Concl}.

\section{Limitations of Tetraquark Pictures}
\label{sec:Limitations}

Each of the major structural pictures described above for the exotic
states presents some dynamical or phenomenological difficulty.
Suppose first that the observed states are something other than true
tet\-ra\-quark resonances.  In this context, we have already
identified the limitations of the hybrid picture.  The cusp effect
does produce phenomena that resemble resonant peaks, but the narrow
$X(3872)$ width ($\Gamma < 1.2$~MeV) appears too small to be
accommodated by a pure cusp not combined with a true
resonance~\cite{Bugg:2008wu}.  Furthermore, the phase motion measured
for the $Z(4475)$~\cite{Aaij:2014jqa} (as well as the $P_c^+$
states~\cite{Aaij:2015tga}) appears to be consistent with that of a
true resonance.

As for true 4-quark bound states, let us begin with a straw man, a
simple democratic molecule of the four quarks.  In addition to this
entity having all the classical instabilities of a 4-body mechanical
system, such a molecule would instantly segregate into attractive
pairs rather than maintain roughly equal spatial
separations~\cite{Vijande:2007rf}.  The easy access to these
``fall-apart'' channels leads one to consider the remaining pictures:
hadrocharmonium, meson molecules, and diquarks.

The hadrocharmonium picture was developed to explain the strong
coupling of several of the exotics to conventional $J/\psi$ or
$\chi_c$ charmonium states.  However, the couplings of the exotics to
$D^{(*)} \bar D^{(*)}$ appear to be just as important [indeed,
dominant for $X(3872)$].  Moreover, it is unclear why the embedded
$c\bar c$ pair would remain dynamically stable with respect to the
light-quark cloud.

The diquark picture, on the other hand, tends to overpredict the
number of bound states, due to its rich color structure; dynamical
assumptions (such as which spin couplings
dominate~\cite{Maiani:2014aja}) are necessary to reduce the number of
states.  Moreover, the forces assembling the diquarks, while strong,
are still smaller than those between $q\bar q$ pairs into color
singlets.  One would expect diquark molecules with typical interquark
separations to re-segregate into meson molecules.

The meson molecular picture is extremely attractive because of its
simplicity and the remarkable proximity of many exotics to two-hadron
thresholds, as in Eq.~(\ref{eq:massdiffs}).  However, several of the
exotics lie far from such thresholds [{\it e.g.}, $Z(4475)$], and
others lie slightly {\em above\/} thresholds [{\it e.g.}, $Y(4260)$
about 30~MeV above $m_{D_s^*} + m_{\bar D_s^*}$], casting doubt on
them being bound states.  However, the most difficult problem for the
meson molecular model is that the most-studied exotic, the $X(3872)$,
despite being extremely weakly bound if it is indeed a molecule
[again, see Eq.~(\ref{eq:massdiffs})], is produced in large amounts
({\it prompt production}) in high-energy colliders such as the
LHC~\cite{Chatrchyan:2013cld}.  In such experiments, quarks are nearly
never created with sufficiently small $p_\perp$ to form a state as
delicately bound as $X(3872)$; and while final-state interactions
({\it i.e.}, $\pi$ exchanges between $D^0$ and $\bar D^{*0}$)
substantially expand the $p_\perp$ range that allows molecules to
form~\cite{Artoisenet:2009wk,Artoisenet:2010uu}, they do not seem to
be enough to explain $X(3872)$ production~\cite{Esposito:2013ada,
  Guerrieri:2014gfa}.

\section{The Dynamical Diquark Picture}
\label{sec:BHLpicture}

The dynamical diquark picture of Ref.~\cite{Brodsky:2014xia} is
motivated by several interesting features of the charmoniumlike
exotics.  As discussed above, the static diquark picture seems
problematic due to the possibility of immediate recombination into
meson molecules.  In order to avoid this recombination, we propose
that the diquark-antidiquark ($\delta$-$\bar \delta$) pair, once
created, rapidly separate to distances $r$ at which the overlaps with
meson wave functions become small.  That is, hadronization occurs
through the exponentially suppressed large-$r$ tails of the mesonic
wave functions, in which $\delta$ supplies the quarks and $\bar
\delta$ the antiquarks.  Since each diquark carries color, the
$\delta$-$\bar \delta$ pair cannot separate indefinitely, but rather
convert their kinetic energy into the potential energy of a color flux
tube stretching between them.

\begin{figure}[ht]
\centering
\includegraphics[height=3in]{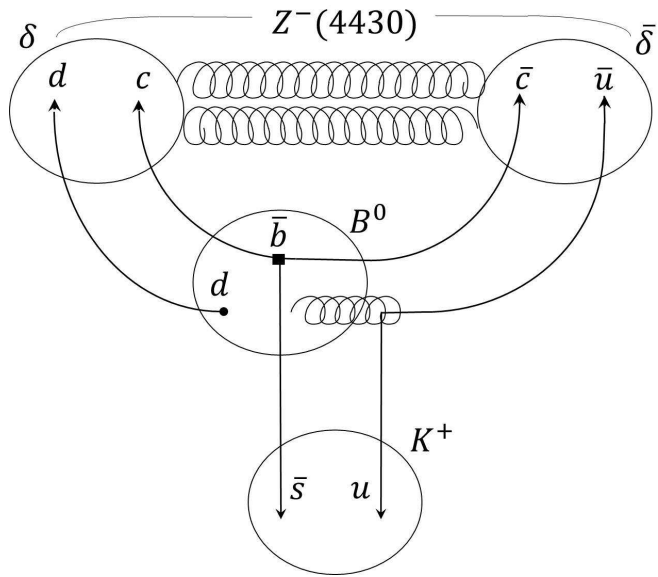}
\caption{Production mechanism for the $Z^-(4475)$ [from $B^0 \to
  Z^-(4475) K^+$], where $\delta$ and $\bar \delta$ indicate the
  diquarks.  The black square indicates the weak decay $\bar b \to
  \bar c c \bar s$.}
\label{fig:Z4475}
\end{figure}

In Fig.~\ref{fig:Z4475} we exhibit the proposed production mechanism
for the $Z^-(4475)$ [previously called $Z^-(4430)$] in the process
$B^0 \to Z^-(4475) K^+$.  One immediately sees that the tetraquark
state in this picture is an entirely new type of bound state: not a
molecule whose components occupy well-defined orbits, but a dynamical
object whose diquark components separate and are distinguishable only
due to the large initial kinetic energy imparted to them (here, via
$\bar b \to \bar c c \bar s$).  One then sees that either path to
hadronization, to ($D^{(*)} + \bar D^{(*)}$) or to (charmonium + light
meson), requires both mesons to have a spatial wave function extent at
least as large as the separation between the diquarks.  The wave
function suppressions lead to a suppressed transition amplitude, and
hence a suppressed observable width.

Of course, if a $B^0$ meson decays via $\bar b \to \bar c c \bar s$,
the simplest decay modes to construct are two-body modes like
$D^{(*)-} D_s^{(*)+}$.  Even then, due to the large phase space
available, each two-body channel accounts for only about 1\%, or
collectively for about 5\%, of the total decay
rate~\cite{Agashe:2014kda}; in the typical $B^0$ $\bar c c \bar s$
decay, multiple hadrons are produced.  The two-body charmonium decay
modes, such as $J/\! \psi \, K^{*0}$, individually occur at the
$10^{-3}$ level, and collectively around 2.3\%.  The branching
fractions into $X(3872)$ or $Z^-(4475)$, occurring at levels ranging
up to $1.7 \cdot 10^{-4}$ for exclusive channels, are at very
reasonable levels for states with a $\delta$-$\bar \delta$ origin,
considering the somewhat smaller attraction within $\delta$ or $\bar
\delta$ compared to that within a color-singlet meson.

Some elements in a picture like Fig.~\ref{fig:Z4475} are familiar from
textbook discussions of confinement.  When the components of a
confined state are produced with a large relative momentum (as in
jets), they stretch a flux tube between them until the color field
contains enough energy to form hadrons by creating an additional
$q\bar q$ pair from the vacuum ({\it string fragmentation}).  In the
picture of Fig.~\ref{fig:Z4475}, the fragmentation occurs as soon as
the threshold for creation of the lightest baryon pair, $\Lambda_c^+
\bar \Lambda_c^{\, -}$ ($2m_{\Lambda_c} = 4573$~MeV), is passed.
Indeed, the lightest exotic above this threshold, $X(4630)$, is
observed to decay predominantly into $\Lambda_c^+ \bar \Lambda_c^{\,
  -}$, exactly as expected in this picture.

In principle, hadronization can occur at any point during the
$\delta$-$\bar \delta$ separation.  However, the standard WKB
semiclassical approximation predicts the transition probability to be
maximal near the classical turning point, {\it i.e.}, when the kinetic
energy of the $\delta$-$\bar \delta$ pair converts entirely into the
potential energy of the flux tube.  The question then becomes how far
apart the $\delta$ and $\bar \delta$ separate before coming to rest.
To estimate this distance $r$, we use that $\delta$ and $\bar \delta$
are somewhat compact (by virtue of each containing a heavy $c$ quark)
color-triplet states and employ the famous linear-plus-Coulomb Cornell
potential~\cite{Eichten:1978tg}, which has been quite
successful~\cite{Barnes:2005pb} in explaining the conventional
charmonium spectrum:
\begin{equation} \label{eq:Cornell} V(r) = -\frac 4 3
  \frac{\alpha_s}{r} + b r + \frac{32\pi\alpha_s} {9m_\delta^2} \left(
    \frac{\sigma}{\sqrt{\pi}} \right)^3 \! \!  e^{-\sigma^2 r^2} {\bf
    S}_{\delta^{\vphantom I}} \!  \cdot {\bf S}_{\bar \delta} \,
  ,
\end{equation}
where $\alpha_s = 0.5461$, $b = 0.1425$~GeV$^2$, $\sigma =
1.0946$~GeV, and $-4/3$ is the color factor specific to {\bf 3}-$\bar
{\bf 3}$ attraction.  The $\delta$ mass can be estimated from lattice
or QCD sum rule calculations, but in any case it is close to the $D$
meson mass, and similarly for its charge radius ($\sim 0.4$~fm).
Applied to the $Z(4475)$, this procedure gives $r = 1.16$~fm, and
$0.56$~fm for $X(3872)$.  In comparison, Eq.~(\ref{eq:Cornell})
applied to charmonium gives $\langle r_{J/\psi} \rangle = 0.39$~fm and
$\langle r_{\psi(2S)} \rangle = 0.80$~fm.  One therefore has a simple
explanation of a remarkable experimental fact: The $Z(4475)$ decays to
$\psi(2S)$ at least 10 times more frequently than to
$J/\psi$~\cite{Chilikin:2014bkk}, despite both states having the same
quantum numbers and much greater phase space available to the
$J/\psi$: It is a simple matter of the large $\delta$-$\bar \delta$
state having a greater wave function overlap with $\psi(2S)$ than
with $J/\psi$.

\section{Applications}
\label{sec:Applications}

\subsection{Dynamical Diquark Resonances and the Cusp Effect}

Why should the dynamical diquark picture produce resonant states, and
why should these states lie close to the meson-meson thresholds?
Three clues noted above lead to a partial answer to this question:
First, the $\delta$-$\bar \delta$ states should have observably small
widths ({\it e.g.}, compare $\Gamma[Z(4475)] = 180 \pm 31$~MeV to
$\Gamma[\rho(770)] = 150$~MeV).  Second, the $\delta$ diquarks and $D$
mesons are expected to have similar masses, so the corresponding
threshold masses are similar.  Third, the cusp effect can combine with
existing resonances to drag the observed resonant mass toward
meson-meson thresholds~\cite{Bugg:2008wu}.  The latter point,
especially, is examined in Ref.~\cite{Blitz:2015nra}, where it is seen
that each such threshold can easily drag a resonant mass several MeV
towards it (even, possibly, overshooting the threshold and creating a
slightly above-threshold resonance).  In the case of the $X(3872)$,
Eq.~(\ref{eq:massdiffs}) shows that several thresholds are clustered
together, potentially creating a strong compound cusp that attracts a
resonance to this mass.

\begin{figure}[ht]
\centering
\includegraphics[height=3in]{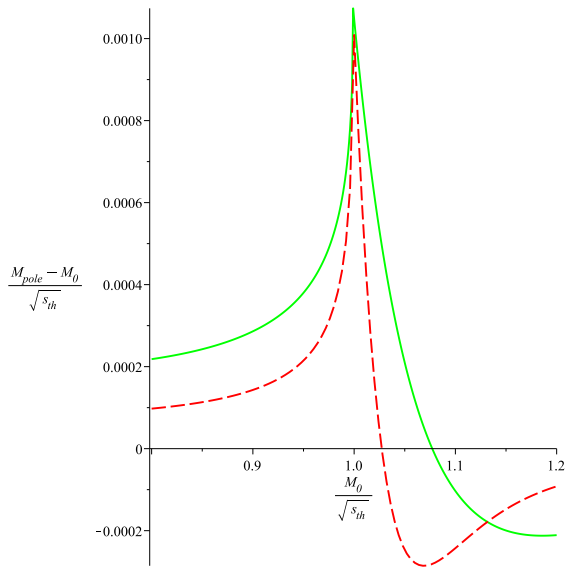}
\caption{Comparison of the effectiveness of resonant pole dragging by
cusps as a function of $M_0/ \! \sqrt{s_{{\rm th},i}}$, from the
diquark cusp (solid, green) and from the mesonic cusp (dashed, red).
Here, $\sqrt{s_{{\rm th},i}} = 3.872$~GeV, $M_0$ and $M_{\rm pole}$
are the bare and final positions of the resonance pole masses, and
couplings have been scaled to give the diquark and meson cusp
functions the same height.}
\label{fig:Cusps}
\end{figure}

Indeed, inasmuch as the diquarks in the $\delta$-$\bar \delta$ state
can separate substantial distances before being forced to hadronize
despite being confined, they approach the behavior of asymptotically
free states and can generate a threshold cusp of their own.  Since the
form factor for creating a multiquark state is expected to fall off at
a different rate than that for mesons due to QCD {\it constituent
  counting rules\/} (first discussed in
Refs.~\cite{Brodsky:1973kr,Matveev:1973ra}), the diquark cusp often
turns out to be broader and more effective at attracting resonant
poles than the meson cusp (see Fig.~\ref{fig:Cusps}).  The full
detailed spectrum of the charmonium sector may turn out to be due to a
rich combination of ``bare'' resonances with locations jostled about
by both meson-meson and $\delta$-$\bar \delta$ thresholds.

\subsection{Exotics and Constituent Counting Rules}

The constituent counting rules are obtained from the twist dimension
of the interpolating fields that refer to the hadrons at short
distances.  They determine the Mandelstam $s$ power-law dependence of
cross sections and form factors for processes at large $s$ and fixed
scattering angle $\theta_{\rm cm}$: The power of $s$ is determined by
the total number $n$ of fundamental constituents (incoming plus
outgoing) appearing in the hard scattering.  In essence, they amount
to counting the number of large-energy propagators necessary to effect
the finite-angle scattering of all the constituents.  In particular,
the invariant amplitude ${\cal M}$ for such a process scales
as~\cite{Brodsky:2015wza}
\begin{equation} \label{eq:constituentcount}
{\cal M} \propto \frac{1}{s^{\frac{n}{2} - 2}} \, .
\end{equation}
From Eq.~(\ref{eq:constituentcount}), the electromagnetic form factor
of a charged tetraquark state such as $Z_c^+ = Z(4475)$ is seen to
scale at large $s$ as
\begin{equation} \label{eq:ffactorcount}
F_{Z_c^+} (s) \to \frac 1 {s^{\frac{1}{2} (1+1+4+4) - 2}} =
\frac{1}{s^3} \, .
\end{equation}
This result is used, {\it e.g.}, in creating Fig.~\ref{fig:Cusps}.
Here we have taken the natural expectation of 4 fundamental
constituents for a tetraquark state.  However, if the $Z_c^+$ contains
diquarks that are so tightly bound that they act as fundamental units
in high-energy scattering processes, then one expects $F_{Z_c^+} (s)
\to 1/s^1$.

While the scaling rules strictly hold only for large $s$ (presumably
several GeV above production threshold), their reach may be extended
to lower energies by taking the ratios of cross sections of processes
that differ primarily through the number of fundamental constituent
components, thus eliminating systematic corrections common to both
processes.  As an example, the ratio
\begin{equation}
\frac{\sigma( e^+ e^- \to Z^+_c(\bar c c \bar d u) + \pi^-(\bar u d))}
{\sigma(e^+ e^- \to \mu^+ \mu^-)} = \left| F_{Z_c, \pi} (s) \right|^2
\propto \frac{1}{s^{n-4}} \, ,
\end{equation}
scales as $1/s^4$ if $Z^+_c$ acts as a two-quark, two-antiquark bound
state, while if the diquarks are particularly tightly bound and act as
fundamental constituents in the hard scattering, the scaling drops to
$1/s^2$.  Similarly, consider the ratio
\begin{equation} \label{eq:Z_Lambdac}
\frac{ \sigma (e^+ e^- \to Z^+_c(\bar c c \bar d u) + \pi^-(\bar u d)
)}
{\sigma (e^+ e^- \to \Lambda_c(cud) \bar{\Lambda}_c (\bar c \, \bar u
\bar d) ) } \propto \frac{1}{s^0} \, ,
\end{equation}
such that the same number of constituents, as well as the same
heavy-quark ($\bar c c$) constituents, appear in both processes.  In
this case, not only the high-$s$ scaling but also corrections due to
the heavy-quark mass cancel in the ratio.  One expects the absolute
numerical value of the ratio to be substantially smaller if $Z_c^+$
behaves as a meson-meson molecule than a $\delta$-$\bar
\delta$ state since the color forces in the former are of the residual
van der Waals type and hence much weaker.

\subsection{The $P_c^+$ Pentaquark Candidates}

Suppose, with reference to Fig.~\ref{fig:Z4475}, that one replaces the
$B^0$ meson with the baryon $\bar \Lambda_b$, so that the $d$ quark is
replaced by the quark pair $\overline{ud}$~\cite{Lebed:2015tna}.
Indeed, let us take the charge conjugate of the diagram so that one
may work with baryons rather than antibaryons.  The light-quark pair
in any $\Lambda_Q$ baryon ($Q = s, c, b$) has, since the earliest days
of QCD, been considered to be a diquark, because it is easily seen to
be a ``good'' spin-0, isoscalar, color-$\bar{\bf 3}$ combination.
The presence of the heavy quark $b$ tends to confine the $(ud)$ to a
small space (one estimate for the root-mean square matter radius of
$\Lambda_b$ is 0.22~fm~\cite{Albertus:2003sx}), so the diquark
$\delta^\prime = (ud)$ can be considered compact, acting essentially
as a spectator in the same way as the $d$ in Fig.~\ref{fig:Z4475}.

The work in Ref.~\cite{Brodsky:2015wza} also argued that one can build
up more complicated multiquark states (pentaquarks, hexaquarks, {\it
etc.})---many of which would be easily produced at facilities such as
the upgraded JLab---by exploiting the attraction of sequentially
formed color-{\bf 3} and -$\bar {\bf 3}$ combinations.  In the case
under discussion here, the $\delta^\prime$ can combine with a
color-$\bar {\bf 3}$ $\bar c$ to form a compact color-{\bf 3} {\it
antitriquark\/} $\theta = \bar c (ud)$.  The same reasoning as that in
Sec.~\ref{sec:BHLpicture} surrounding Fig.~\ref{fig:Z4475} then shows
that a {\it pentaquark\/} with valence quark structure $c\bar c uud$,
the result of a rapidly separating color-{\bf 3} antitriquark $\theta
= \bar c (ud)$ and color-$\bar{\bf 3}$ diquark $\delta = (cu)$, is an
absolutely natural result of the picture.  It is the central point of
Ref.~\cite{Lebed:2015tna} that this description explains the newly
discovered states~\cite{Aaij:2015tga} $P_c^+(4380)$ and $P_c^+(4450)$.

As with the diquark-antidiquark picture for tetraquark states, the new
diquark-triquark picture has a static
antecedent~\cite{Karliner:2003dt}, which was used to explain the
then-extant pentaquark candidate $\Theta^+ (1535) = u \bar s udd$.
However, that model also made use of a color-{\bf 6} diquark inside
the triquark in order to achieve a desirable level of binding energy
with respect to the nearby $KN$ threshold.

\section{Conclusions}
\label{sec:Concl}

We have proposed an entirely new dynamical picture for understanding
the exotic charmoniumlike states, such as $X(3872)$, $Z(4475)$,
$P_c^+(4380)$, and $P_c^+(4450)$ that have been discovered in recent
years and are still being uncovered today.  We propose that at least
some subset of them are bound, but not molecular, states of color-{\bf
3} and color-$\bar{\bf 3}$ compact diquarks and triquarks, which have
achieved substantial separation due to the large energy release of the
process in which they are formed.  The states remain bound only due to
the confinement of these colored components, and can only decay when
color-singlet combinations form through the large-$r$ tails of wave
functions of mesons or baryons stretching from one colored component
to the other.

Such states can be studied through their multiquark nature using
constituent counting rules, as well as their potential ability to
create threshold cusps.  The next stage of investigation will be to
explore the dynamics of the diquark and triquark formation and the
mechanism (going beyond simple quantum-mechanical ideas) by which
hadronization across the flux tube is accomplished.  The question of
why such states occur at some masses and not others, producing the
rich spectroscopy already observed, must also be addressed beyond the
confines of static Hamiltonian models.  Nevertheless, the dynamical
picture already gives tantalizing hints of what phenomenology might be
possible, in an energy regime that until recently was thought to be a
very well-understood sector of particle physics.

\Acknowledgements This work would not have been possible without the
dedicated efforts of my collaborators, Stan Brodsky, Dae Sung Hwang,
and Sam Blitz.


\end{document}